\documentclass[twocolumn,showpacs,amsmath,amssymb,prl,floatfix,superscriptaddress]{revtex4}
\usepackage{graphicx}% Include figure files
\usepackage{dcolumn}% Align table columns on decimal point
\usepackage{bm}% bold math
\usepackage{bbm}
\usepackage{multirow,amssymb,amsbsy,amsmath}
\usepackage{stmaryrd}
\newcommand{\ddim}{\udelta\kern0.1em}

\newcommand{\beikonst}[2]{\left( #1 \right)_{\kern-0.2em #2}}

\newcommand*{\bra}[1]{\mathopen{\langle}#1\mathclose{|}}
\newcommand*{\ket}[1]{\mathopen{|}#1\mathclose{\rangle}}
\newcommand{\braket}[2]{\mathopen{\langle}#1\mathclose{|}#2\mathclose{\rangle}}

\newcommand{\ketbra}[1]{\mathopen{|}#1\mathclose{\rangle}\hspace{-0.25em}\mathopen{\langle}#1\mathclose{|}}
\newcommand{\ketbrap}[2]{\mathopen{|}#1\mathclose{\rangle}\hspace{-0.25em}\mathopen{\langle}#2\mathclose{|}}

\begin{document}
%\linenumbers

%\preprint{APS/123-QED}

% -----------------------------------------
%
% Title
%

\title{Collectively Enhanced Interactions in Solid-state Spin Qubits}

\author{Hendrik Weimer}%
\email{hweimer@cfa.harvard.edu}%
\affiliation{Physics Department, Harvard University, 17 Oxford Street, Cambridge, MA 02138, USA} %
\affiliation{ITAMP, Harvard-Smithsonian Center for Astrophysics, 60 Garden Street, Cambridge, MA 02138, USA}%
\author{Norman Y. Yao} %
\affiliation{Physics Department, Harvard University, 17 Oxford Street, Cambridge, MA 02138, USA} %
\author{Mikhail D. Lukin} %
\affiliation{Physics Department, Harvard University, 17 Oxford Street, Cambridge, MA 02138, USA}

\date{\today}% 

\begin{abstract}

We propose and analyze a technique to collectively enhance interactions between solid-state quantum registers composed from random networks of spin qubits. In such systems, disordered dipolar interactions generically result in localization. Here, we demonstrate the emergence of a single collective delocalized eigenmode as one turns on a transverse magnetic field.  The interaction strength between this symmetric collective mode and a remote spin qubit is  enhanced by square root of the number of spins participating in the delocalized mode. 
Mediated by such collective enhancement, long-range quantum logic between remote spin registers can occur at distances consistent with optical addressing. A specific implementation utilizing Nitrogen-Vacancy defects in diamond is discussed and the effects of decoherence are considered. 

\end{abstract}

% PACS, the Physics and Astronomy Classification Scheme.

\maketitle

Harnessing collective phenomena by utilizing ensembles of identical
particles is a powerful tool, which has been exploited in effects
ranging from superradiance to scattering suppression
\cite{Inouye1999}. The coherent dynamics resulting from interactions
with individual constituents of an ensemble are often too weak to be
observed directly; however, as evidenced by experiments in systems
such as Rydberg atoms \cite{Heidemann2007,Gaetan2009,Urban2009},
cavity QED \cite{Brennecke2007,Colombe2007}, atomic ensembles
\cite{Bartenstein2004,Simon2007}, and solid state qubits
\cite{Zhu2011}, collective enhancement provides a natural route to
overcoming this challenge. Here, we demonstrate that, for electronic
spin quantum registers, such collective effects enable an extended
coherent coupling over large distances --- an essential prerequisite
for quantum information processing.

Owing to favorable coherence properties, electronic spins associated
with point-like defects in solid-state systems have garnered significant
recent interest as candidates for room-temperature quantum
registers. Quantum control of such spins can be achieved using a
combination of optical, magnetic and electric fields. While our considerations apply
to a variety of electronic spin qubits \cite{Koehl2011,Pla2012,Chanier2012}, here, we focus on the
Nitrogen-Vacancy (NV) center in diamond.  The NV center harbors an
 electronic spin ($S=1$), which can be optically initialized, coherently
manipulated and read out on sub-wavelength scales
\cite{Jelezko2004,Childress2006, Dutt2007}. These results have sparked
several recent proposals which utilize networks of NV registers as the
platform for a scalable quantum information processor
\cite{Yao2011,Bermudez2011,Weimer2012,Yao2012}.  However, for any
spin qubit candidate, two crucial challenges remain to be addressed:
1) the weakness of the magnetic dipolar interactions on distances
compatible with individual optical addressing and 2) the disorder in
spin positioning due to inherent imperfections during defect creation.

\begin{figure}[tb]
\includegraphics[width=0.8\linewidth]{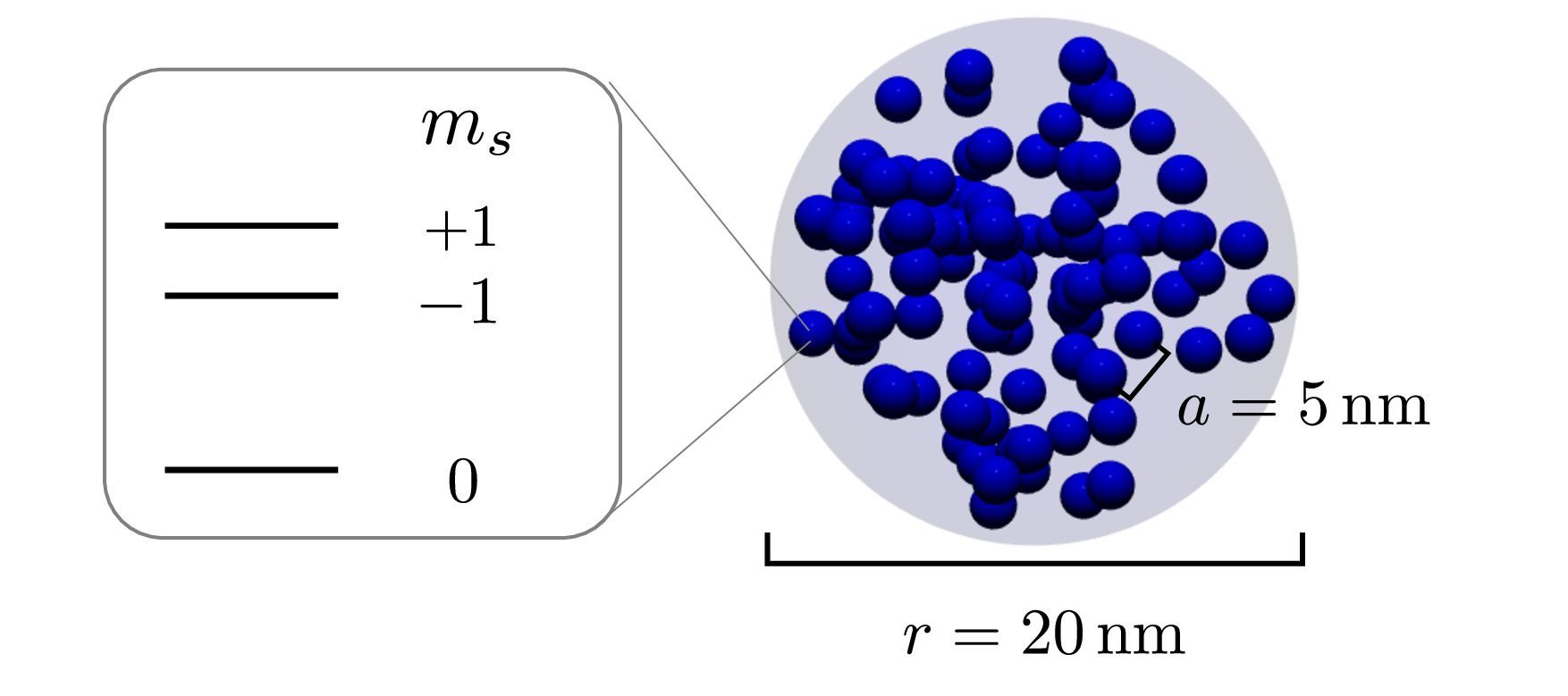}
\caption{High-density NV spin ensemble distributed randomly within a
  sphere of diameter $r$, with an average distance $a$. The NV centers
  have three internal spin states that are split by a zero-field
  splitting and a Zeeman field.}
\label{fig:setup}
\end{figure}
 
In this Letter, we present a novel approach to remote quantum logic
which harnesses collectively enhanced interactions to overcome both of
the above challenges.  The key idea underlying our proposal is to
associate a single, robust qubit with a collective, generally
disordered spin-ensemble (Fig.~\ref{fig:setup}).  If the spins behave
in an aggregate fashion, such a qubit can produce a large
state-dependent magnetic field, leading to enhanced long-range
coupling between ensembles; this is reminiscent of tailored
light-matter interactions achieved via atomic ensembles
\cite{Sherson2006}. However, we note that quenched disorder naturally
leads to localization in solid-state spin systems \cite{Mirlin1996}
due to random flip-flop interactions. Similar to Anderson localization
\cite{Anderson1958}, this implies that each eigenmode of the ensemble
is composed of only a few spins. Here, we demonstrate the use of a
uniform transverse magnetic field to overcome this issue.  The applied
field causes the symmetric $W$-state \cite{Jaksch2000,Lukin2001} to
become an approximate eigenstate of the Hamiltonian, thereby enabling
us to harness it as a collective qubit. Moreover, we show that this
particular state is largely insensitive to the underlying spin
distribution and hence robust to effects of disorder.

To be specific, we now describe our proposal in the context of NV
diamond color centers. The largest energy scale in this system
($\Delta$) is set by a combination of the zero field splitting
($2.87\,\mathrm{GHz})$ and the projection of the external Zeeman field
along the NV axis. However, we would like to stress that the alignment
of the field with the NV axis is not crucial, as its quantization axis
is essentially given by the zero-field splitting. We assume that the
Zeeman field is sufficiently strong to ensure that the $m_s = -1$ spin
state is sufficiently far detuned and hence does not contribute to the
effective dynamics.  Thus, the number of $m_s=1$ spins, $m$, is an
approximately good quantum number and a perturbative description is
justified. The second-largest energy scale arises due to the
perturbation created by the transverse field $\Omega$. To gain a
qualitative understanding, let us restrict ourselves to the
analytically tractable case where $m$ is either $0$ or $1$. The
effective Hamiltonian is, $H_r = -\Delta \ketbra{} +
\sqrt{N}\Omega(\ketbrap{0}{W}+\mathrm{h.c.})$, where the state
$\ket{0}$ has all ensemble spins polarized into $m_s=0$, and the
collective $\ket{W}$ state is fully symmetric with all spins sharing a
single excitation,
\begin{equation}
\ket{W} = \frac{1}{\sqrt{N}}\sum\limits_i\ket{0\ldots 1_i \ldots}.
\end{equation}
Second order perturbation theory in $\sqrt{N}\Omega/\Delta$ yields
$H_r' = -(\Delta+J)\ketbra{\tilde{0}}+J\ketbra{\tilde{W}}$, with
$J=N\Omega^2/\Delta$ and the tilde referring to the perturbed eigenstates. Including higher $m$ manifolds  merely leads
to a renormalization of $J$, without changing this qualitative picture 
(so long as we are in the perturbative limit). This is equally true in
the presence of dipolar interactions, provided that the energy scale
$J$ is larger than the characteristic  strength of the dipolar
interaction $V_{dd}$. Thus, even with these additional terms,  the new eigenstates will still have
substantial overlap with the collective $\ket{W}$ state. This is in
stark contrast to the situation without a transverse field, where 
strongly quenched disorder owing to random spin positions localizes
all such eigenstates, even in three dimensions. Furthermore, the
dipolar interaction naturally ensures that collective states with
different $m$ values will have different energies, leading to a
``blockade''-type scenario, where manifolds with $m>1$ are
energetically inaccessible \cite{Jaksch2000,Lukin2001}. This allows us to
selectively drive the transition between $\ket{0}$ and $\ket{W}$
without populating any other collective states, provided that the
external driving $\Omega_\mathrm{ext}$ is weaker than $V_{dd}$. This hierarchy of energy scales can be summarized as: $\Delta \gg J
\gg V_{dd} \gg \Omega_\mathrm{ext}$.

Let us consider a three dimensional ensemble of $N=100$ NV centers randomly
 distributed within a diameter $r=20\,\mathrm{nm}$,
as depicted in 
Fig.~\ref{fig:setup}. Such high density NV ensembles have been recently realized using long-time annealing of repeat-electron-irradiated diamond samples
\cite{Toyli2010,Spinicelli2011,Hausmann2011,Isoya2012}. 
We will characterize our effective two-level system ($m_s = 0,1$) using Pauli spin operators $\sigma_\alpha$. Being magnetic dipoles, NV centers interact with one another via long-range magnetic dipolar interactions (ignoring energy non-conserving terms which are suppressed by the NV center's zero field splitting),
\begin{eqnarray}
    V_{ij} &=& \left(1-3\cos^2\vartheta_{ij}\right)\frac{\mu^2}{|{\bf r}_i -{\bf r}_j|^3}\\
    &\times&
\left\{\frac{1}{4}\left[1+\sigma_z^{(i)}\right]\left[1+\sigma_z^{(j)}\right] - \sigma_+^{(i)}\sigma_-^{(j)}-\sigma_-^{(i)}\sigma_+^{(j)}\right\},\nonumber
\end{eqnarray}
where ${\bf r}_i$ denotes the position, $\mu$ characterizes the magnetic
dipole moment, and $\vartheta_{ij}$ is the angle between the
NV axis and the vector connecting sites ${\bf r}_i$ and ${\bf r}_j$. 
 The total Hamiltonian including both on-site and interaction terms is
then given by $H = \Delta/2\sum_i\sigma_z^{(i)} + \Omega \sum_i\sigma_x^{(i)} + \sum_{i<j}V_{ij}$.

Now, let us turn to the enhanced coupling between an isolated NV defect (hereon termed ``qubit'') and the collective ensemble, separated by the distance $R$. We envision the ensemble to be initialized into the $\ket{0}$ state, while the NV qubit is initialized to the $m_s=1$
state. By ensuring that the qubit splitting is tuned resonant with only the  $\ket{W}$ state, one finds that the effective dynamics are  restricted to the single-excitation manifold of
the combined qubit-ensemble system; to lowest order, these dynamics are governed by,
\begin{equation}
  H_\mathrm{eff} = \sqrt{N_c}\frac{\mu^2}{R^3} (\ketbrap{1_q,0}{0_q,W} + \mathrm{h.c.}),
\label{eq:heff}
\end{equation}
where $N_c$ characterizes the approximate number of spins
participating in the $\ket{W}$ state and the notation $\ket{1_q,0}$
refers to the combined state with the NV qubit being in $m_s=1$ and
with the ensemble spins being in $\ket{0}$. Consistent with
sub-wavelength techniques such as stimulated emission depletion (STED)
microscopy ($R=100\,\mathrm{nm}$), we will assume that the NV qubit
can be manipulated and read out independently of the ensemble
\cite{Maurer2010}.

\begin{figure}[t]
%  \psfrag{N}[c][c]{$N_c$}
%  \psfrag{E}[c][c]{$E [h\times\mathrm{GHz}]$}
%  \psfrag{ 0}[c][c]{$0$}
%  \psfrag{ 20}[c][c]{$20$}
%  \psfrag{ 40}[c][c]{$40$}
%  \psfrag{ 60}[c][c]{$60$}
%  \psfrag{ 80}[c][c]{$80$}
%  \psfrag{ 1}[c][c]{$1$}
%  \psfrag{-2}[c][c]{$-2$}
%  \psfrag{-1}[c][c]{$-1$}
%  \psfrag{ 2}[c][c]{$2$}
%\includegraphics[width=\linewidth]{fig/dressingE}
\includegraphics[width=\linewidth]{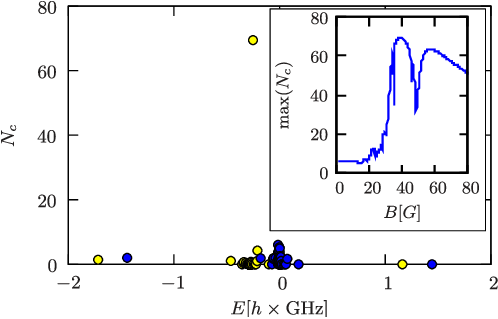}
\caption{Comparison of the collective enhancement $N_c$ for $B=0$
  (blue) and $B = 40\,\mathrm{G}$ (yellow) within the
  single excitation manifold. In the latter case there is a
  collectively enhanced state with $N_c \approx 70$, corresponding to
  an increase by more than one order of magnitude ($\Delta = h\times
  4\,\mathrm{GHz})$. The inset shows the maximum value of $N_c$
  depending on the transverse field strength $B$.}
\label{fig:nc}
\end{figure}

To support the qualitative picture presented above, we now perform
exact diagonalization of the interacting spin Hamiltonian. In the
majority of the numerics, we restrict ourselves to $m \leq2$
excitations; however, we check the validity of our results by
including the $m=3$ manifold for slightly smaller system sizes. For
each eigenstate $\ket{\phi}$, we calculate the collective enhancement
factor, defined as
\begin{equation}
  N_c = \left(\sum\limits_i^N \braket{0_1\ldots 1_i \ldots 0_N}{\phi}\right)^2,
\end{equation}
which for a symmetric eigenmode characterizes the
number of participating ensemble spins. As expected,
in the absence of a transverse field, disorder localizes all
eigenstates and as depicted in Fig.~\ref{fig:nc} (blue circles), $N_c
\ll N$ for all eigenstates. On the other hand, In the case of a
moderate transverse field $\Omega=\mu B$, with $B \approx 40\,\mathrm{G}$, one finds the existence of a single eigenstate
with $N_c \approx 70 \sim N$.  While the specific details of this
state depend on the microscopic details (e.g., spin distribution
within the ensemble, higher-order couplings to the
  $m_s = -1$ state, and magnitude of the applied transverse field),
its collective nature is rather robust. In particular, as one varies
the strength of the transverse field $B$, there exists a large
parameter regime where $N_c >50$
(Fig.~\ref{fig:nc}). This result clearly supports our
  previous analytical arguments on the existence of a symmetric
  collective mode. The dips in $N_c$ are associated with resonance
effects, which arise when other eigenstates become near-degenerate
with the collective state.  Finally, the decrease of $N_c$ for large
values of $\Omega$ signals the breakdown of perturbation theory as
$\sqrt{N}\Omega/\Delta$ approaches unity.

\begin{figure}[b]
%\psfrag{ 0}[c][c]{$0$}
%\psfrag{ 0.5}[c][c]{$0.5$}
%\psfrag{ 1.5}[c][c]{$1,5$}
%\psfrag{ 2}[c][c]{$2$}
%\psfrag{ 2.5}[c][c]{$2.5$}
%\psfrag{ 3}[c][c]{$3$}
%\psfrag{ 0.2}[c][c]{$0.2$}
%\psfrag{ 0.4}[c][c]{$0.4$}
%\psfrag{ 0.6}[c][c]{$0.6$}
%\psfrag{ 0.8}[c][c]{$0.8$}
%\psfrag{ 1}[c][c]{$1$}
%\psfrag{p}[c][c]{$p_q$}
%\psfrag{t}[c][c]{$t [\mathrm{ms}]$}
%\includegraphics{fig/rabi}
\includegraphics[width=\linewidth]{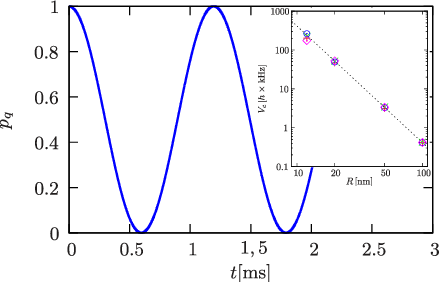}
\caption{Collectively enhanced Rabi oscillations between an isolated
  NV qubit and a NV ensemble. The probability to find the qubit in the
  $m_s=1$ state, $p_q$, goes to zero within a time $t_\pi \approx
  600\,\mathrm{\mu s}$. The inset shows the collectively enhanced
  coupling strength $V_c$ between the qubit and the ensemble for four
  different realizations. The dashed line shows the asymptotic $1/R^3$
  dependence.}
\label{fig:rabi}
\end{figure}

We now perform simulations of the combined
qubit-ensemble system. As previously discussed, the system is
initialized to $\ket{1_q,0}$ and the qubit splitting is tuned resonant
with the energy of the collective mode; the resulting dynamics is
evinced in Fig.~\ref{fig:rabi}. Interestingly, the probability of finding
the qubit in the $m_s=1$ state, $p_q$, exhibits collectively enhanced
Rabi oscillations.  The frequency of these oscillations is enhanced by
nearly an order of magnitude relative to that expected for bare
dipolar interactions between two individual NV qubits at a similar
distance.  The numerics also allow us to obtain the time required for
an interaction-induced $\pi$ pulse, $t_\pi$ and from this, one can
derive the effective distance $R$ associated with
$H_\mathrm{eff}$. Surprisingly, in all cases, we observe that this
distance corresponds not to $R-r$, but instead to the distance between
the NV qubit and the \emph{center} of the ensemble. We study the
effects of putting the qubit closer to the ensemble by calculating the
collectively enhanced coupling strength $V_c$ (as  extracted from the numerically obtained $t_\pi$). As shown in
Fig.~\ref{fig:rabi}, we find that only for distances very close to the
ensemble does the collective enhancement deviates from the asymptotic
$1/R^3$ scaling, e.g., the qubit is coupled to individual spins rather
than to the entire ensemble.

%\begin{figure}[tb]
%  \psfrag{N}[c][c]{{$\max(N_c)$}}
%  \psfrag{g}[c][c]{{$B [G]$}}
%  \psfrag{ 0}[c][c]{{$0$}}
%  \psfrag{ 20}[c][c]{$20$}
%  \psfrag{ 40}[c][c]{{$40$}}
%  \psfrag{ 60}[c][c]{$60$}
%  \psfrag{ 80}[c][c]{{$80$}}
%  \psfrag{ 50}[c][c]{$50$}
%  \psfrag{ 100}[c][c]{{\Large $100$}}
%  \psfrag{ 150}[c][c]{$150$}
%  \psfrag{ 200}[c][c]{{\Large $200$}}
%  \psfrag{ 250}[c][c]{$250$}
%\includegraphics{fig/dressing3d}
%\includegraphics{fig3}
%\caption{Maximum value of the collective enhancement $N_c$ depending
%  on the transverse field strength $\Omega$.}
%\label{fig:Omega}
%\end{figure}

%\begin{figure}[b]
%\psfrag{ 0}[c][c]{{\large $0$}}
%\psfrag{ 0.25}[c][c]{{\large $0.25$}}
%\psfrag{ 0.5}[c][c]{{\large $0.5$}}
%\psfrag{ 0.75}[c][c]{{\large $0.75$}}
%\psfrag{ 1}[c][c]{{\large $1$}}
%\psfrag{ 25}[c][c]{{\large $25$}}
%\psfrag{ 50}[c][c]{{\large $50$}}
%\psfrag{ 75}[c][c]{{\large $75$}}
%\psfrag{ 100}[c][c]{{\large $100$}}
%\psfrag{N}[c][c]{{\large $N_c(R)/N_c(\infty)$}}
%\psfrag{R}[c][c]{{\large $R\,[\mathrm{nm}]$}}
%\includegraphics{fig/R}
%
%\end{figure}

%\begin{figure}[b]
%\psfrag{ 0.1}[c][c]{{\large $0.1$}}
%\psfrag{ 1}[c][c]{{\large $1$}}
%\psfrag{ 10}[c][c]{{\large $10$}}
%\psfrag{ 20}[c][c]{{\large $20$}}
%\psfrag{ 50}[c][c]{{\large $50$}}
%\psfrag{ 100}[c][c]{{\large $100$}}
%\psfrag{ 1000}[c][c]{{\large $1000$}}
%\psfrag{V}[c][c]{{\large $V_c\,[h\times\mathrm{kHz}]$}}
%\psfrag{R}[c][c]{{\large $R\,[\mathrm{nm}]$}}
%\includegraphics{fig/R2}
%
%\end{figure}

\emph{Experimental Realization and Decoherence.---} Thus far, our
discussion has assumed that both the NV qubit and the ensemble spins
are perfectly decoupled from the environment. In any experimentally
realistic scenario, however, there are two natural error sources
which will be present: spin decoherence and spin
relaxation.  We are particularly interested in the
scaling of the error rates with $N$, as this may adversely
affect the scaling fidelity of our proposed long-range gates
\cite{Yao2012}. As the error processes act locally on individual
spins, we first calculate the error rate for a single spin and
multiply the result by $N$ to obtain the rate for the collective
state. For simplicity, we assume that the collective state is the
previously described $\ket{W}$ state in which a single excitation is
shared among all $N$ spins.

First, let us consider the effects of spin
decoherence. The worst-case
scenario for such decoherence is given by the leaking out  into non-symmetric states.  Consequently, the error
probability after a single $T_2$ decoherence event on spin $i$ is given
by the probability to leave the $\ket{W}$ state,
\begin{equation}
  p_{\bar{W}} = p_{T_2}\left[1-|\bra{W}\sigma_z^{(i)}\ket{W}|^2\right] = \frac{4}{N} p_{T_2}\left(1-\frac{1}{N}\right),
\end{equation}
where $p_{T_2}$ is the single spin decoherence rate.  For large $N$,
this result is essentially independent of $N$ (after weighing with the
number of spins); therefore, the effect of $T_2$ processes on such a
collective $\ket{W}$ state does not get enhanced by system size and in
fact, is only slightly worse than for a single spin, i.e., it can be
expressed in terms of an effective coherence time
$T_2^\mathrm{eff}$.

Second, we consider the errors  arising from phonon-induced spin relaxation
processes ($T_1$). Here, we must distinguish between processes which flip an
ensemble spin from $m_s = 1$ to $m_s=0$, and the reverse. This
asymmetry can  easily be seen by noting that the $\ket{W}$ state has
only one spin in $m_s=1$, while all other spins are in $m_s=0$. We
denote the error probability associated with these two events as $p_{T_1}^{1\to
  0}$ and $p_{T_1}^{0\to 1}$, respectively. For $p_{T_1}^{1\to
  0}$, the
state $\ket{0}$ with all ensemble spins in $m_s=0$ is not affected at
all, while the probability to flip from the $\ket{W}$ state into
$\ket{0}$ is given by
\begin{equation}
  p_{W\to 0} = p_{T_1^{1\to 0}}|\bra{0}\sigma_-^{(i)}\ket{W}|^2 = \frac{p_{T_1^{1\to 0}}}{N},
\end{equation}
which is again independent of the size of the ensemble after rescaling
with $N$. 

However, this is not the case for $T_1^{0 \to 1}$
processes. Both the $\ket{0}$ and the $\ket{W}$ state are strongly
affected by such processes, since the existence of any additional spin in the $m_s =1$
state corresponds to an effective magnetic impurity; this impurity modifies the energy of the collective
state, thus tuning it out of resonance with the NV
qubit. Additionally, this new state is also no longer an eigenstate of the
Hamiltonian; numerical simulations demonstrate that this state dephases very
quickly due to dipolar interactions within the
ensemble. Thus, since any single spin $T_1^{0 \to 1}$ error will immediately decohere the collective state,  the effective error rate owing to $p_{T_1}^{0\to 1}$ is enhanced by $N$ and scales with the size of the ensemble.

While the system size scaling of $p_{T_1}^{0\to 1}$ errors might seem
unfortunate, in solid-state spin systems, it is often the case that
$T_1 \gg T_2$. Our proposed protocol is particularly useful in cases
where $T_1/N$ remains longer than $T_2$, implying that the ensemble's
noise is dominated by decoherence as opposed to the enhanced
relaxation. The specific example of NV centers highlights this
crucial point. The decoherence of the NV originates from fluctuating
magnetic fields as neighboring pairs of dipoles flip-flop
\cite{Balasubramanian2009,Bar-Gill2012}. Even at low temperatures it
is impossible to freeze out such magnetic fluctuations and $T_2$
remains on the order of milliseconds
\cite{Balasubramanian2009,Maurer2012}. On the other hand, the
relaxation of the NV is thought to originate from an Orbach
spin-phonon process; such a process has an \emph{exponential}
dependence on temperature and implies that even moderate cooling can
yield exceedingly long $T_1$ times ($\gg 1$s at cryogenic
temperatures) \cite{Redman1991,Harrison2006,Takahashi2008}. By liquid
nitrogen temperatures, the errors introduced by the enhanced $T_1$
processes are already sub-percent, enabling us to focus on the effects
of decoherence. An alternate approach to combat the enhanced
relaxation of the collective state is to utilize conventional
dynamical decoupling techniques (e.g., WAHUHA) \cite{Waugh1968} to
suppress dipolar interactions within the individual ensembles.

\begin{figure}[tb]
\includegraphics[width=\linewidth]{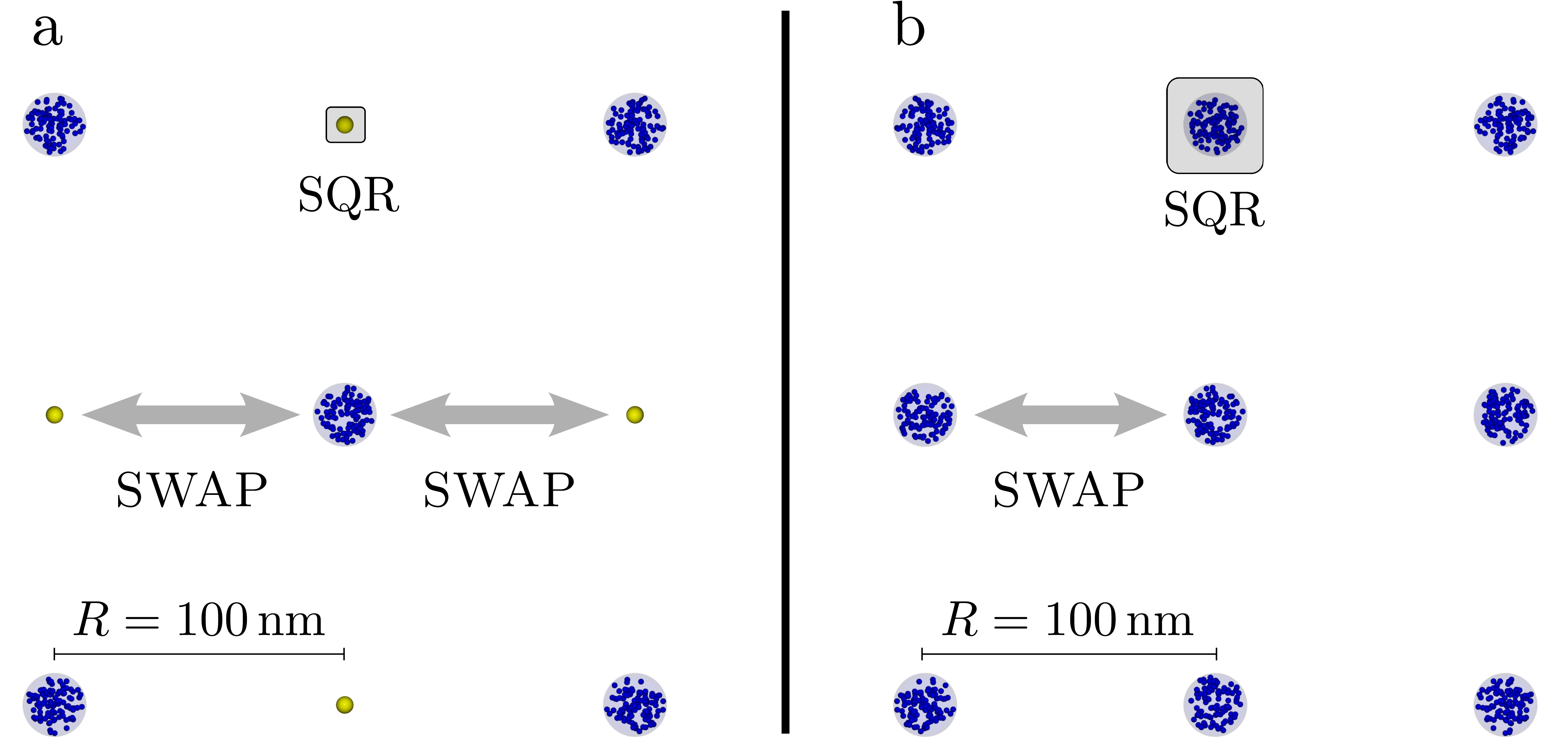}
\caption{Scalable architectures with collectively enhanced
  interactions, corresponding to a lattice spacing of
  $R=100\,\mathrm{nm}$, compatible with sub-wavelength optical
  addressing. (a) Individually addressable NV qubits (yellow) are used
  for single qubit operations (SQR), whereas the collectively enhanced
  interaction with an ensemble is used to mediate two-qubit gates via
  {\sc SWAP} operations. (b) NV ensembles are used as collective
  qubits, where also single qubit operations are performed using the
  collective $\ket{W}$ state.}
\label{fig:arch}
\end{figure}

\emph{Collective quantum gates.---} We now turn to a possible
application where isolated NV qubits are interspersed with
high-density NV ensembles, forming a regular structure, as depicted in
Fig.~\ref{fig:arch}a. The qubits are used for initialization,
single-qubit rotations, and readout. Two-qubit gates between remote
spin qubits are mediated by the ensemble between them and thus benefit
from collectively enhanced interactions. We would like
  to point out that such architectures put only modest requirements on
  the positioning of the NV centers; in particular, the positional
  disorder within the NV ensembles is essentially irrelevant. The
gate time $t_g$ is limited by the {\sc SWAP} time $t_\pi$ required to
to transfer the information from one of the qubits to the ensemble
(required four times per gate operation ) \cite{Yao2011a}. The
resulting error (assuming $T_1/N \gg T_2$) of the gate is given by
$\varepsilon = 1-\exp[-(4 t_\pi/ T_2^\mathrm{eff})^3]$ in the presence
of spin echo decoupling \cite{Maze2008}. For an error of $\varepsilon
= 10^{-2}$, this translates to a required coherence time of
$T_2^\mathrm{eff} = 11\,\mathrm{ms}$, which can be readily realized in
isotopically pure diamond samples
\cite{Balasubramanian2009,Maurer2012} or by using dynamical decoupling
pulses \cite{deLange2010,Bar-Gill2012}. The requirements on the
coherence time can be further relaxed by increasing the number of
spins in the ensemble or by reducing the qubit-ensemble separation.

An architecture featuring even better gate fidelities can be realized
using a collective encoding scheme for the qubits (see
Fig.~\ref{fig:arch}b). There, the logical $\ket{0}$ state corresponds
to all nuclear spins being polarized, while the logical $\ket{1}$
state is a collective nuclear spin $\ket{W}$ state. This state can be
prepared by applying a microwave pulse to map the electronic $\ket{W}$
state onto a nuclear spin $\ket{W}$ state \cite{Lukin2001}. The
timescale for such a single qubit operation is limited to
approximately $100\,\mathrm{kHz}$ by the hyperfine splitting of the NV
centers in the $m_s=1$ state ($A_{\parallel} \approx
-2.14\,\mathrm{MHz}$ for $^{14}\mathrm{N}$)
\cite{Felton2009}. Operation at cryogenic temperatures
  allows for resonant read-out of the $\ket{W}$ state via the zero
  phonon line without significant background fluorescence
  \cite{Robledo2011}. In this collective qubit architecture,
two-qubit gates between ensembles are enhanced by a factor of $N$
instead of $\sqrt{N}$, thus leading to a {\sc SWAP} time of $t_\pi =
70\,\mathrm{\mu s}$.  Thus, we find that a gate error of $\varepsilon
= 10^{-2}$ requires a coherence time of
$T_2^\mathrm{eff} = 700\,\mathrm{\mu s}$, while for $\varepsilon =
10^{-4}$, a coherence time of $T_2^\mathrm{eff} =
3\,\mathrm{ms}$ is needed
\cite{Balasubramanian2009,Maurer2012,Ryan2010,Naydenov2011}.

In summary, we have shown that collectively enhanced interactions can
be realized between an NV qubit and a mesoscopic NV ensemble. Our
proposed approach relies upon a transverse magnetic field to inhibit
the localization of symmetric $W$-eigenstate. Our work enables the
realization of collectively enhanced quantum gates with high fidelity
and provides an important step towards the realization of scalable
quantum information architectures involving solid-state electronic
spins.

\begin{acknowledgments}

We acknowledge fruitful discussions with C.~Laumann and
S.~Bennett. This work was supported by the National Science Foundation
through a grant for the Institute for Theoretical Atomic, Molecular
and Optical Physics at Harvard University and Smithsonian
Astrophysical Observatory, a fellowship within the Postdoc Program of
the German Academic Exchange Service (DAAD), the DOE (FG02-
97ER25308), CUA, NSF, DARPA, AFOSR
MURI, and the Packard Foundation.

\end{acknowledgments}

%\bibliographystyle{aip}
%\bibliography{/home/hendrik/da/hendrik}
%\bibliography{/home/itp/weimer/hendrik}
%\bibliography{blob}
%\bibliography{/Users/nyao/Dropbox/Shared-NY-HW/the blob}

\end{document}